\documentclass[onecolumn,showpacs,preprintnumbers,amsmath,amssymb]
{revtex4}

\usepackage{epsfig}

\begin{document}

\title{Exact time-average distribution for a stationary
non-Markovian massive Brownian particle coupled to two heat baths}

\author{D. O. Soares-Pinto}
 \email{dosp@cbpf.br}
\affiliation{CICECO, Universidade de Aveiro, 3810-193, Aveiro,
Portugal} \affiliation{Centro Brasileiro de Pesquisas
F\'{\i}sicas, Rua Dr. Xavier Sigaud 150, CEP 22290-180, Rio de
Janeiro, Brazil}
\author{W. A. M. Morgado}
 \email{welles@fis.puc-rio.br}
\affiliation{Departamento de F\'{\i}sica, Pontif\'{\i}cia
Universidade Cat\'olica do Rio de Janeiro,  22452-970 Rio de
Janeiro, Brazil}

\date{\today}

\begin{abstract}
Using a time-averaging technique we obtain exactly the probability
distribution for position and velocity of a Brownian particle
under the influence of two heat baths at different temperatures.
These baths are expressed by a white noise term, representing the
fast dynamics, and  a colored noise term, representing the slow
dynamics. Our exact solution scheme accounts for inertial effects,
that are not present in approaches that assume the Brownian
particle in the over\,--\,damped limit. We are also able to obtain
the contribution associated with the fast noise that are usually
neglected by other approaches.
\end{abstract}


\maketitle

\section{Introduction}

The role of time-scales is crucial for understanding the validity
of equilibrium techniques to any physical process since fast
degrees of freedom (time-scale $\tau_f$) generally reach their
stationary distribution well within the experiment's duration but
the same cannot be assumed about slow ones (time-scale $\tau_s$).
For instance, glasses are a very important class of systems
presenting slow dynamics~\cite{RMP_76_2004_785}.

The probability distribution for the slow degrees of freedom
depends on the system's preparation and also on the elapsed time
interval, i.e., for how long the system has
aged~\cite{RMP_76_2004_785}. For the time interval $\tau_f\ll t
\ll \tau_s$, the system behaves as driven by an effective external
slow field associated with the slow variables. However, there is
no physical reason for the slow and fast fluctuations distribution
to match, so, in general, $T_{slow}\neq T_{fast}$. In this case,
the reduced form for the fast variable's distribution is an
instantaneous equilibrium one~\cite{1986_PhysA_138_231}. For
longer times, $t\gg \tau_s$, the distributions for all variables
(fast and slow) reach a stationary state. For closed systems
governed by a microscopic Hamiltonian, the stationary state will
be a true equilibrium one. For open systems in contact with a
single thermal reservoir at temperature $T$, the final state will
also be the equilibrium distribution. However, by subjecting a
system to simultaneous contacts with two (or more) reservoirs at
distinct temperatures, say $T_1$ and $T_2$, it will reach a
stationary state distinct from the Boltzmann-Gibbs (BG)
equilibrium~\cite{2000_JPSJ_69_247}. In consequence, the
stationary state will depend on the properties of system-reservoir
interactions and the reservoirs temperatures. Simple models
presenting these characteristics, such as Brownian Particles (BP),
have been used to explore the physics of slow dynamics and
glasses~\cite{2000_PRE_62_845, 2007_PLA_360_552, 2006_PLA_357_275,
2005_PRE_72_011112, 2004_PRE_69_016119, 2005_JSTAT_09_P09013,
1999_PhysA_263_242, 2006_JPhysCS_40_76, 2004_PRE_70_046104,
2003_PRE_67_016102}.

The complete description of BP is given by its microscopic
interactions with a heat bath of lighter particles as well as the
interactions with the external environment. Trying to evaluate the
macroscopic behavior starting up from a microscopic model presents
an impossible task. However, for many physical systems we can
often reduce the number of important variables down to a
manageable set, thanks to well separated
time-scales~\cite{sandri1,sandri2, livro_chapman} allowing us to
eliminate the fast variables of the problem. The effect of the
eliminated variables is taken into account by means of a random
forcing term on the equations of motion for the remaining
variables~\cite{1971_JChemPhys_54_3541, 1986_PhysA_138_231} and by
suitable friction coefficients. The random term (noise function)
represents the effect of the collisions of the BP with the heat
baths's particles. In the case these thermal forces vary on a very
short time-scale, their effect can be represented by uncorrelated
collisions (white noise) and the BP motion can be analyzed in a
simple way because its stochastic dynamics can be treated as a
Markovian process. Finally, its long term behavior is well
described by a Boltzmann-Gibbs (BG)
equilibrium~\cite{livro_vankampen}. However, for many realistic
Brownian processes in dense fluids the short time-scale
approximation for the random forces may not be accurate, since the
time-intervals of the microscopic collisions might overlap.
Instead, we need to consider that those forces act upon the BP
during a well defined time-scale, or colored noise. Also, the
presence of excited large-scale hydrodynamic modes will affect the
BP by long time-scale feed-back loops~\cite{1970_PhysRevA_1_18,
1970_JChemPhys_53_3813} that can be accounted for by a dissipative
memory function term in the Langevin-like equation describing the
Brownian motion~\cite{livro_vankampen}.

Our present goal is to obtain the long-time stationary state for a
BP system, subjected to contacts with two different heat baths at
distinct temperatures, that is simple enough to be treated exactly
but sophisticated enough to present the non-trivial stationary
properties discussed above~\cite{2007_arXiv_0705.1951,
livro_nelson}.

In our model, it is possible to keep the BP from reaching the BG
equilibrium by simultaneously subjecting it to the influence of
two thermal baths at different temperatures $T_1$ and $T_2$, one
acts as heat source while the other acts as a heat sink. As for
the BP simultaneously subjected to $T_1$ and $T_2$, the stationary
state will depend on the details of the interaction between the
baths and the BP and the effective temperature will be an
intermediary one between $T_1$ and $T_2$, as will be shown. Even
if temperature gradients are not present, the stationary
distribution for velocities and positions will differ from the BG
measure, reducing to it only when the temperatures are the same,
$T_1=T_2=T$. This simple model has been proposed to describe the
behavior of glasses~\cite{2000_EPJB_16_317,
kurchan_nature,2000_PRE_62_845, 2007_PLA_360_552,
2006_PLA_357_275, 2005_PRE_72_011112, 2004_PRE_69_016119,
2005_JSTAT_09_P09013, 1999_PhysA_263_242, 2006_JPhysCS_40_76,
2004_PRE_70_046104, 2003_PRE_67_016102} where one assumes
different time-scales for each thermal bath, via their noise
properties: the short time-scale white noise reproduces thermal
relaxation whereas the long time-scale noise reproduces very slow,
correlated, structural rearrangement of the glass. Additionally,
in our approach we can include the  effects related to the finite
mass of the particle in stationary state, as well as other effects
associated with the slow noise~\cite{2000_JPSJ_69_247} which are
often neglected by other methods based on the over-damped limit
calculations (which corresponds to the limit $m\rightarrow 0$).
For instance, the BP adjusts itself to the slow noise function as
~\cite{2000_EPJB_16_317, 2000_JPSJ_69_247}, specially when its
mass can be neglected (over-damped case), and the position
distribution becomes very similar to a true equilibrium form,
given that the slow noise acts like an external effective
potential~\cite{2000_JPSJ_69_247}. However, this simplification
eliminates the contribution of the fast noise for the stationary
distribution~\cite{2000_JPSJ_69_247}. In contrast, our
time-averaging scheme will integrate that small effect for a very
long time and exhibit the missing contribution.

This paper is organized as follows. In Section II we describe the
model. In Section III we calculate the time-averaging integrals
obtaining the displacement and velocity non-vanishing terms. In
Sections IV and V, we obtain the stationary (or equilibrium)
distribution and discuss our main results.

\section{Exactly solvable model}

\subsection{Langevin-like Equation}
Let's consider a massive Brownian Particle (BP) moving under the
action of a confining potential $V(x)$, which is in contact with
two heat baths of distinct temperatures and time-scales, expressed
by the noise functions $\xi(t)$ and $\eta(t)$. There are some
interesting physical systems that present this kind of multiple
time-scale phenomena. For instance, in glasses the short
time-scale (fast) modes are associated with thermal vibrations and
the long time-scale (slow) ones with structural reorganization of
the atomic structure~\cite{2000_EPJB_16_317}.

The effect of the fast modes on the BP are represented by the
white noise term $\eta(t)$, associated with the heat bath at
temperature $T_1$ and the instantaneous friction coefficient
$\Gamma_1$. On the other hand, the slow modes contribution are
represented by a colored noise term $\xi(t)$, associated with the
heat bath at temperature $T_2$ (in general $T_2\neq T_1$). These
modes give rise to long-time feed-back loops that are expressed by
the integration of a frictional memory function $\phi(t-t^{'})$
over time.

This is modeled by two coupled stochastic differential equations:
\begin{equation}\label{eq.01}
\dot x=v,
\end{equation}
\begin{equation}\label{eq.02}
m\,\dot v= -\frac{\partial V}{\partial x} - \int_{0}^{t}
dt^{'}\phi(t-t^{'})v(t^{'}) - \Gamma_{1}v(t) +\xi(t) +\eta(t),
\end{equation}
where the properties of the terms above are discussed in the
following. We assume $V(x)= \frac12 kx^2$ for simplicity.

\subsection{Noise properties}
The stochastic process described by Eq.(\ref{eq.02}) is
non-Markovian due to the noise $\xi$ be time correlated with a
well defined time-scale $\tau$ and exponential time-correlation
function, or a colored noise (see for example
\cite{1995_AdvChemPhys_89_239}). It represents the slow dynamics
and only at time-scales much larger than $\tau$ will $\xi$
manifest Markovian behavior. The noise $\eta$ is a white one and
represents the fast dynamics. Both noises, fast and slow, are
Gaussian and can be defined in terms of their two lowest
cumulants. Let's start with the slow noise properties:

\begin{equation}\label{eq.03}
\langle\xi(t)\rangle = 0,
\end{equation}
\begin{equation}\label{eq.04}
\langle\xi(t)\xi(t^{'})\rangle = \frac{D_{2}}{\tau}
\exp\left(-\frac{|t-t^{'}|}{\tau}\right)=\frac{\Gamma_{2}\,T_{2}}{\tau}
\exp\left(-\frac{|t-t^{'}|}{\tau}\right),
\end{equation}
giving the memory function~\cite{2006_PhysA_365_289}
\begin{equation}\label{eq.05}
\phi(t-t^{'})=\frac{\langle\xi(t)\xi(t^{'})\rangle}{T_{2}}
=\frac{\Gamma_{2}}{\tau}\exp\left(-\frac{|t-t^{'}|} {\tau}\right),
\end{equation}
where $\tau$ is the characteristic time-scale of the slow noise,
$T_2$ its temperature and $\Gamma_2$ a dissipative strength
associated with $\xi$. The units are chosen so that Boltzmann's
constant is equal to one ($k_{B}=1$).

The first two cumulants for the fast noise are given below:
\begin{equation}\label{eq.06}
\langle\eta(t)\rangle = 0,
\end{equation}
\begin{equation}\label{eq.07}
\langle\eta(t)\eta(t^{'})\rangle = T_{1}\,\Gamma_{1}\,
\delta(t-t'),
\end{equation}
where $T_1$ is the temperature and $\Gamma_1$ is the dissipative
coefficient for the fast noise.

\subsection{Equilibrium distribution}
Our goal is to find out the exact stationary solution for
Eq.(\ref{eq.02}) through the time-averaging of
\[
\rho(x,v,t)=\delta(x-x(x_o,v_o,t))\delta(v-v(x_o,v_o,t)),
\]
as $t\rightarrow\infty$. The functions $x(x_o,v_o,t))$ and
$v(x_o,v_o,t)$ are the solutions for a given set of BP's initial
conditions $(x_o,v_o)$ and for a given realization of the
stochastic processes $\xi(t)$ and
$\eta(t)$~\cite{livro_vankampen}. We need to take the average over
these last two stochastic processes in order to obtain the
stationary solution.

Thus, the stationary distribution for the Brownian degrees of
freedom reads~\cite{2006_PhysA_365_289}:
\begin{eqnarray}\label{eq.08}
P^{ss}(x,v)&=&\lim_{T\rightarrow \infty}\frac{1}{T}
\int_{0}^{T}dt\,\left<\rho(x,v,t)\right>_{\xi,\eta}\nonumber\\
&=&\lim_{z\rightarrow 0^{+}}
z\int_{0}^{\infty}dt\,e^{-zt}\left<\rho(x,v,t)\right>_{\xi,\eta}
\end{eqnarray}

For simplicity, we will assume that the initial conditions are
$x_o=v_o=0$, since the terms depending on the initial conditions
tend to zero as $t\, \rightarrow\,
\infty$~\cite{2006_PhysA_365_289}.

\subsection{Laplace transformations}
The centerpiece of our method is to use Laplace transforms in
order to make our problem manageable and the mathematical reason
for doing so is because Eq.(\ref{eq.08}) itself is the Laplace
transformation of the averaged density. Thus, our goal is to
describe the stationary state probability distribution,
$P^{ss}(x,v)$, in terms of the Laplace transform of the noise
functions, $\tilde{\xi}(z)$ and
$\tilde{\eta}(z)$~\cite{2006_PhysA_365_289}.

In fact, we can express $P^{ss}(x,v)$ as a sum of many terms
involving $\tilde{\xi}(z)$ and $\tilde{\eta}(z)$. Most of these
terms will not contribute to the distribution, but we can
calculate analytically the remaining ones, finding the exact
expression for $P^{ss}(x,v)$. In order to do that we need to
express the Laplace transforms, of the position and velocity of
the BP, as functions of $\tilde{\xi}(z)$ and $\tilde{\eta}(z)$. In
the following we proceed to implement the strategy just described.

We start by taking the Laplace transformations of
Eqs.(\ref{eq.01}) and (\ref{eq.02}) (with Re$(z)>0$):
\begin{equation}\label{eq.11}
\left[m\,z+\frac{\Gamma_{2}}{\tau\,
z+1}+\Gamma_{1}+\frac{k}{z}\right]\,
\tilde{v}(z)=\tilde{\xi}(z)+\tilde{\eta}(z),
\end{equation}
and
\begin{equation}\label{eq.12}
z\,\tilde{x}(z)=\tilde{v}(z).
\end{equation}

The Laplace transforms for the noise correlations are given
by~\footnote{Calculations are done using Maple 7.0}:

\begin{eqnarray}\label{eq.13}
\langle\tilde{\xi}(iq_i+\epsilon)\,\tilde{\xi}(iq_j+\epsilon)\rangle
&=&
\left\{\frac{2\,\Gamma_{2}\,T_{2}}{\left[i(q_i+q_j)+2\epsilon\right]
\left[1-\tau(iq_i+\epsilon)\right]
\left[1-\tau(iq_j+\epsilon)\right]}\right\}+\\
&-&\Gamma_{2}\,T_{2}\,\tau\left\{\frac{3+\tau\left[i(q_i+q_j)+2\epsilon\right]
-\tau^2(iq_i+\epsilon)(iq_j+\epsilon)}{
\left[1-\tau(iq_i+\epsilon)\right]\left[1+\tau(iq_i+\epsilon)\right]
\left[1-\tau(iq_j+\epsilon)\right]
\left[1+\tau(iq_j+\epsilon)\right]}\right\}\nonumber,
\end{eqnarray}

\begin{equation}\label{eq.14}
\langle\tilde{\eta}(iq_{i}+\epsilon)\,\tilde{\eta}(iq_{j}+\epsilon)\rangle
= \frac{2\,\Gamma_{1}\,T_{1}}{i(q_i+q_j)+2\epsilon},
\end{equation}
\begin{equation}\label{eq.14b}
\langle\tilde{\eta}(iq_{i}+\epsilon)\,\tilde{\xi}(iq_{i}+\epsilon)\rangle
= 0.
\end{equation}
It can be shown that the second member of the RHS of
Eq.(\ref{eq.13}) does not contribute to the stationary
distribution $P^{ss}(x,v)$ when $z\rightarrow 0$ in
Eq.(\ref{eq.08}). The above Eqs.(\ref{eq.13}), (\ref{eq.14}) and
(\ref{eq.14b}) will be crucial for calculating $P^{ss}(x,v)$ via
Eq.(\ref{eq.08}).

We observe that when $\Gamma_{2}\rightarrow 0$ we recover previous
results~\cite{2006_PhysA_365_289} since the contribution of the
slow noise disappears. Furthermore, we notice that by taking the
limit $\tau\rightarrow 0$ the colored noise $\xi$ becomes a white
one, which is reflected in the fact that its contribution in
Eqs.(\ref{eq.02}) and (\ref{eq.11}) becomes similar to the one by
$\eta$.

Following a strategy similar to the used in
Ref.~\cite{2006_PhysA_365_289}, we express the Laplace
transformation of position and velocity as functions of the
Laplace transformation of the noise functions:
\begin{equation}\label{eq.15}
\tilde{v}(z)=\Theta(z)\left[\,\tilde{\xi}(z)+\tilde{\eta}(z)\,\right],
\end{equation}
\begin{equation}\label{eq.16}
\tilde{x}(z)=\Omega(z)\left[\,\tilde{\xi}(z)+\tilde{\eta}(z)\,\right],
\end{equation}
with
\begin{equation}\label{eq.17}
\Theta(z)= z\, \Omega(z)=\frac{z\,(\tau\, z+1)}{m\,\tau
\left(z^3+\frac{\theta}{m\,\tau}\,z^2+\frac{\omega}{m\,
\tau}\,z+\frac{k}{m\,\tau}\right)},
\end{equation}
where $\theta$ and $\omega$ are given by
\begin{equation}\label{eq.18}
\theta\equiv m+\Gamma_{1}\,\tau,
\end{equation}
and
\begin{equation}\label{eq.19}
\omega\equiv \Gamma_{1}+\Gamma_{2}+k\,\tau.
\end{equation}

For simplicity, we assumed the initial conditions $x(t=0)=0$ and
$v(t=0)=0$, since for $T\rightarrow \infty$ the terms carrying the
memory of the initial conditions will decay to
zero~\cite{2006_PhysA_365_289}. The denominator of $\Theta(z)$ and
$\Omega(z)$ is of the third order on $z$, thus, there are three
distinct roots for:
\begin{equation}\label{eq.20}
z^3+\frac{\theta}{m\,\tau}\,z^2+\frac{\omega}{m\,\tau}\,z+\frac{k}{m\,\tau}=0,
\end{equation}
namely $z_1$, $z_2$ and $z_3$:
\begin{equation}\label{eq.21}
z_{1}=-\frac{(\theta\,\lambda+\sigma
-\lambda^2)}{3\,m\,\tau\,\lambda},
\end{equation}
\begin{equation}\label{eq.22}
z_{2}=z_3^{*}=-\frac{(2\,\theta\,\lambda-\sigma
+\lambda^2)}{6\,m\,\tau\,\lambda}+i\frac{\sqrt{3}\,(\sigma
+\lambda^2)}{6\,m\,\tau\,\lambda},
\end{equation}
where
\begin{equation}\label{eq.23}
\sigma\equiv -\theta^{2}+3\,\omega\,m\,\tau,
\end{equation}
and
\begin{equation}\label{eq.24}
\lambda\equiv \frac{1}{2^{1/3}} \sqrt [3]{9\,\omega\,\theta\,m\tau
-27\,k \,{m\tau }^{2}-2\,{\theta}^{3}+3\,m\tau\,[3\,(4\,{\omega}^{
3}m\tau -{\omega}^{2}{\theta}^{2}-18\,\omega\,\theta\, m\tau
\,k+27\,{k}^{2}{m\tau }^{2}+4\,k\,{\theta}^{3})]^{1/2} }.
\end{equation}

We define the quantities $\mathcal{A}$, $\mathcal{B}$,
$\mathcal{C}$, $\mathcal{D}$, $\mathcal{F}$ and $\mathcal{G}$ as:
\begin{equation}\label{eq.25}
z_{2}\equiv \mathcal{A}+i\,\mathcal{B}\,;\, z_{1}-z_{2}\equiv
\mathcal{C}+i\,\mathcal{D}\,;\,z_{1}+z_{2} \equiv
\mathcal{F}+i\,\mathcal{G}.
\end{equation}

The terms $\mathcal{A}$, $\mathcal{B}$, $\mathcal{C}$,
$\mathcal{D}$, $\mathcal{F}$ and $\mathcal{G}$ are important
because we perform residue integrations over their
poles~\cite{2006_PhysA_365_289}. It is important to compute the
expressions for the leading terms in the expressions
(\ref{eq.23}), (\ref{eq.24}) and (\ref{eq.25}) when
$m\,\tau\,\rightarrow 0$, i.e., in the over\,--\,damped limit.
Since $\theta>0$, in linear order on $m\,\tau$ we have
\begin{eqnarray}
\lambda  &=&
(-\theta)\left(1-\frac{3\,\omega\,m\,\tau}{2\,{\theta}^{2}}
-\frac{m\,\tau\,\sqrt {3}}{2\,{\theta}^{2}} \sqrt
{-{\omega}^{2}+4\,k\,\theta}\right).
\end{eqnarray}
The function $\sigma$ is already linear in $m\,\tau$. In the
leading order in $m\,\tau$, we obtain:
\begin{equation}\label{eq.c}
\mathcal{A} \sim  -\frac{\omega}{2\,\theta},\,  \mathcal{B} \sim
\frac{\sqrt{-\omega^2+4\,\theta\,k}}{2\,\theta},\, \mathcal{C}
\sim -\frac{\theta}{m\,\tau},\, \mathcal{D} \sim
-\frac{\sqrt{-\omega^2+4\,\theta\,k}}{2\,\theta},\, \mathcal{F}
\sim -\frac{\theta}{m\,\tau},\, \mathcal{G} \sim
\frac{\sqrt{-\omega^2+4\,\theta\,k}}{2\,\theta}.
\end{equation}

\subsection{Calculating the stationary distribution}

In Appendix~\ref{appa} we show the derivation for the following
equation (originally obtained in Ref.~\cite{2006_PhysA_365_289})
for the stationary state distribution Eq.~\ref{eq.08}:
\begin{eqnarray}\label{eq.26}
P^{ss}(x,v) &=&\lim_{z,\epsilon\rightarrow0^{+}}
\sum_{l,m=0}^{\infty} \int_{-\infty}^{+\infty}\frac{dQ}{2\pi}
\frac{dP}{2\pi}e^{iQx+iPv} \frac{(-iQ)^{l}}{l!}
\frac{(-iP)^{m}}{m!} \int_{-\infty}^{+\infty} \prod_{f=1}^{l}
\frac{dq_{f}}{2\pi}\prod_{h=1}^{m}\frac{dp_{h}}{2\pi} \nonumber\\
&\times&\frac{z}{z-\left[\sum_{f=1}^{l}iq_{f}+
\sum_{h=1}^{m}ip_{h}+(l+m)\epsilon\right]} \langle
\prod_{f=1}^{l}\tilde{x}(iq_{f}+\epsilon)\prod_{h=1}^{m}
\tilde{v}(ip_{h}+\epsilon) \rangle
\end{eqnarray}

\begin{figure}[ht]
\includegraphics[scale = 0.7, angle = 0]{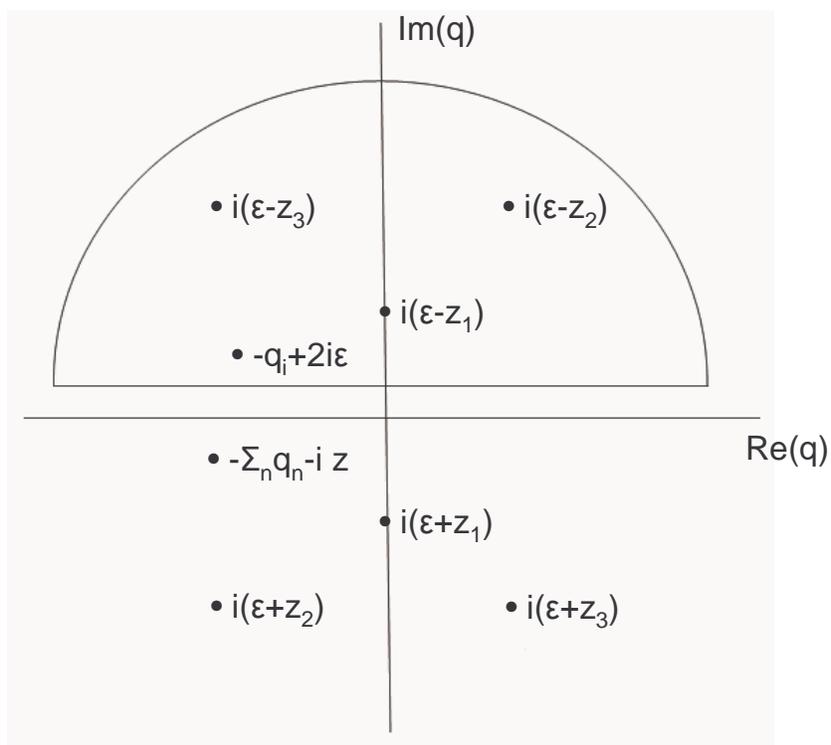}
\caption{Integration path for the q or p--variables.} \label{fig}
\end{figure}

In order to compute $P^{ss}$ we need to replace Eqs.(\ref{eq.13})
to (\ref{eq.16}), into Eq.(\ref{eq.26}). The integrations paths
for the variables $\left\{q_{f},p_{h}\right\}$ can be chosen to be
identical to the one shown in Fig.(\ref{fig}): the integrations in
Eq.(\ref{eq.26}) will correspond to a series of residue
integrations over (only some) of the multiple poles of its
integrand. These poles correspond to the ones associated with the
noise functions (Eqs.(\ref{eq.13}) to (\ref{eq.16})) and the ones
given by Eq.(\ref{eq.17}) and the function $I(z)$:
\begin{equation}\label{Iz}
I(z)=\frac{z}{z-\left[\sum_{f=1}^{l}iq_{f} +
\sum_{h=1}^{m}ip_{h}+(l+m)\epsilon\right]}.
\end{equation}

In special, the function $I(z)$ is crucial for our computations
but its role only becomes clear near the end of the calculations.
As a matter of fact, the pole of I(z) is located outside of the
integration path for all $\left\{q_{f},p_{h}\right\}$, as in
Fig.(\ref{fig}). The really important poles are associated with
the roots of the denominators of $\Theta(z)$ and $\Omega(z)$,
$z_1$, $z_2$ and $z_3$ above (the z-poles), and the ones from the
denominators of the Laplace transforms for the noise variances
(the noise-poles), Eqs.(\ref{eq.13}) and (\ref{eq.14}), that are
locate on the inside of the path of Fig.(\ref{fig}). After
replacing the rather complicated forms for the noise Laplace
transforms into Eq.(\ref{eq.26}), the computing of $P^{ss}(x,v)$
becomes a task of doing the residue integrations around the
relevant poles and collecting the non-vanishing terms.

The effect on $I(z)$ due to the residue integration of
Eq.(\ref{eq.26}) over its relevant poles is rather interesting.
The integrations over the noise-poles for
$\left\{q_{f},p_{h}\right\}$ that are present in $I(z)$ imply that
the value the denominator of $I(z)$ will take has a non-zero real
part, even if only one of such an integration has been done. In
that case, when taking the limit $z\,\rightarrow\,0$, we obtain
$lim_{z\,\rightarrow\,0}I(z) = 0$. However, whenever an
integration over any one of the $\left\{q_{f},p_{h}\right\}$
variables at a noise-pole is done, it has the effect of reducing a
sum of the form $i(q+p)+2\epsilon$, $i(q+q')+2\epsilon$ or
$i(p+p')+2\epsilon$ to zero in the denominator of $I(z)$. It means
that the other member of the pair may be integrated around any of
its poles without contributing to the denominator of $I(z)$,
making that integral a possible non-zero contribution to
$P^{ss}(x,v)$, when the limit $z\,\rightarrow\,0$ is taken. Thus,
we notice that only the residue integrations that eliminate the
$\left\{q_{f},p_{h}\right\}$ variables from the denominator of
$I(z)$, two of them at a time, will reduce $I(z)$ to
\[
lim_{z\,\rightarrow\,0} I(z) = lim_{z\,\rightarrow\,0} \frac{z}{z}
= 1.
\]
Only in this case, $I(z)=1$, the corresponding term will
contribute to the stationary state
distribution~\cite{2006_PhysA_365_289}.

It is straightforward to show that cross terms that couple
$\tilde{x}$ and $\tilde{v}$ ($\Omega\Theta$-type) do not
contribute, while other terms coupling $\tilde{x}$ and $\tilde{x}$
($\Omega\Omega$-type) or $\tilde{v}$ and $\tilde{v}$
($\Theta\Theta$-type) will give non-zero contributions. As a
consequence, the stationary probabilities for positions and
velocities are independent: $P^{ss}(x,v) = P^{ss}(x) P^{ss}(v)$.

In the following, we evaluate some typical integrals and deduce
the stationary distribution.

\section{Time-averaging}

\subsection{Inertial case}

In this section, we evaluate the terms that generate the
stationary state distribution $P^{ss}(x,v)$ by analyzing the
distinct contributions grouped into Eq.(\ref{eq.26}). So, lets
start by studying a pair-integration contribution from a
$\Omega\Omega -$correlation integral (for a position-position
correlation this is a $[q,q']$ pair, such as mentioned in the last
section):

\begin{equation}\label{eq.27}
\int_{-\infty}^{+\infty}\frac{dq_{i}}{2\pi}\frac{dq_{j}}{2\pi}
\frac{z}{z-i(q_{i}+q_{j}-2i\epsilon)-i\diamond}\,\Omega(iq_{i}+\epsilon)\Omega(iq_{j}+\epsilon)
\left[\langle\tilde{\eta}(iq_{i}+\epsilon)\tilde{\eta}(iq_{j}+\epsilon)\rangle+
\langle\tilde{\xi}(iq_{i}+\epsilon)\tilde{\xi}(iq_{j}+\epsilon)\rangle\right]
= \frac{z}{z-i\diamond}\, W_{\Omega},
\end{equation}
where we integrate over the contributing poles (only), and
$i\diamond=\sum_{f=1,\neq i,j}^{l}iq_{f}+
\sum_{h=1}^{m}ip_{h}+(l+m-2)\epsilon$ represents the summation
over the $p$'s and $q$'s that are not integrated over in the
denominator of $I(z)$. The $W_{\Omega}$ factor is given below (see
Appendix~\ref{appc}):
\begin{eqnarray}\label{eq.28}
W_{\Omega}&=&\frac{\Gamma_{1}\,T_{1}}{2\,m^{2}}\,
\frac{2\,z_{1}\,\mathcal{A\,B}+(\mathcal{B\,D\,G-B\,C\,F})+(\mathcal{A\,D\,F+A\,C\,G})}
{\mathcal{A\,B}\,(\mathcal{C}^{2}+\mathcal{D}^{2})\,(\mathcal{F}^{2}+\mathcal{G}^{2})}+\nonumber\\
&&
-\frac{(\Gamma_{1}\,T_{1}+\Gamma_{2}\,T_{2})}{2\,m^{2}\,\tau^{2}}\,
\frac{2\,\mathcal{A\,B}\,(\mathcal{A}^{2}+\mathcal{B}^{2})-z_{1}\,(\mathcal{B\,D\,G-B\,C\,F})+z_{1}\,(\mathcal{A\,D\,F+A\,C\,G})}
{z_{1}\,\mathcal{A\,B}\,(\mathcal{A}^{2}+\mathcal{B}^{2})\,(\mathcal{C}^{2}+\mathcal{D}^{2})\,(\mathcal{F}^{2}+\mathcal{G}^{2})},
\end{eqnarray}
The terms depending on $\Gamma_1\,T_1$ ($\Gamma_2\,T_2$) result
from the integration over the poles of the
$\langle\tilde{\eta}\tilde{\eta}'\rangle$
($\langle\tilde{\xi}\tilde{\xi}'\rangle$), and the effect of the
integration over $q_{i}$ and $q_{j}$ on Eq.(\ref{eq.27})
corresponds to the appearance of a $W_{\Omega}$ factor and the
reduction of the denominator of $I(z)$ from
$\left[z-i(q_{i}+q_{j}-2i\epsilon)-i\diamond\right]$ to
$\left[z-i\diamond\right]$.

Similarly, a typical pair-integration contribution from a
$\Theta\Theta -$correlation integral (velocity-velocity
correlation $\rightarrow\,\,[p,p']$ pair) reads:
\begin{equation}\label{eq.30}
\int_{-\infty}^{+\infty}\frac{dp_{i}}{2\pi}\frac{dp_{j}}{2\pi}
\frac{z}{z-i(p_{i}+p_{j}-2i\epsilon)-i\diamond}\,\Theta(ip_{i}+\epsilon)\Theta(ip_{j}+\epsilon)
\left[\langle\tilde{\eta}(ip_{i}+\epsilon)\tilde{\eta}(ip_{j}+\epsilon)\rangle+
\langle\tilde{\xi}(ip_{i}+\epsilon)\tilde{\xi}(ip_{j}+\epsilon)
\rangle\right]=\frac{z}{z-\diamond}W_{\Theta}
\end{equation}
where the $W_{\Theta}$ factor is
\begin{eqnarray}\label{eq.31}
W_{\Theta}&=&\frac{(\Gamma_{1}\,T_{1}+\Gamma_{2}\,T_{2})}{2\,m^{2}\,\tau^{2}}\,
\frac{2\,z_{1}\,\mathcal{A\,B+(B\,D\,G-B\,C\,F)+(A\,D\,F+A\,C\,G)}}
{\mathcal{A\,B}\,(\mathcal{C}^{2}+\mathcal{D}^{2})\,(\mathcal{F}^{2}+\mathcal{G}^{2})}+ \nonumber\\
&&-\frac{\Gamma_{1}\,T_{1}}{2\,m^{2}}\,\frac{2\,z_{1}^{3}\,\mathcal{A\,B}+
(\mathcal{A}^{3}-3\,\mathcal{A}\,\mathcal{B}^{2})
(\mathcal{C\,G+D\,F})+(\mathcal{B}^{3}-3\,\mathcal{A}^{2}\,\mathcal{B})(\mathcal{C\,F-D\,G})}
{\mathcal{A\,B}\,(\mathcal{C}^{2}+\mathcal{D}^{2})\,(\mathcal{F}^{2}+\mathcal{G}^{2})}.
\end{eqnarray}

It is straightforward to show that the integrals of the type
$\Theta\Omega -$correlation  (for a $[q,p]$ pair) over the correct
poles will not contribute to $P^{ss}(x,v)$. It means that the
information about the position distribution ($q$'s) does not get
mixed up with the information about the velocity distribution
($p$'s). In other words, the distributions are independent
($P^{ss}(x,v) = P^{ss}(x) P^{ss}(v)$) as mentioned before.

Schematically, we can represent the integral in Eq.(\ref{eq.26})
by:
\[
\lim_{z\rightarrow0}\int \prod_{j=1}^{2n}\,dp_j
\prod_{l=1}^{2m}\,dq_l\,
\frac{z}{z-i\diamond}\prod_{ij}^{\mbox{all }p\mbox{-pairs}}
\langle\tilde\chi_i\tilde\chi_j\rangle\prod_{kl}^{\mbox{all
}q\mbox{-pairs}} \langle\tilde\chi_k\tilde\chi_l\rangle\,
\rightarrow \, \left(\begin{array}{c} \mbox{all}\\\mbox{terms}
\end{array}\right)\,W_{\Theta}^n\,W_{\Omega}^m,
\]
where $\tilde\chi$ represents the noise functions' Laplace
transforms. The number of terms above remains to be  evaluated.

First we notice that we can factor out the $W_{\Theta}^n$
contributions from the $W_{\Omega}^m$ contributions. By doing so,
we only need to calculate the number of ways of integrating all
$[q,q']$ pairs (and similarly for the  $[p,p']$ pairs).
Furthermore, averaging over a power $2t$ of $\tilde\xi$'s and
$\tilde\eta$'s yields a number of terms analogous to the number of
ways of distributing $2t$ balls into $t$ boxes that have room for
two balls each. The factor above is given by:
\[
u_{2t}=\frac{(2t)!}{2^t\,t!}.
\]
Let's replace that into the equation for $P^{ss}(x)$. The factored
out contribution for the position distribution reads
\begin{eqnarray}
P^{ss}(x)  &\rightarrow& \sum_{l=0}^{\infty}
\int_{-\infty}^{+\infty}\frac{dQ}{2\pi}e^{iQx}
\frac{(-iQ)^{2l}}{(2l)!}\,\frac{(2l)!}{2^l\,l!}\,W_{\Omega}^{l},
\nonumber
\end{eqnarray}
\begin{eqnarray}\label{eq.29}
\Rightarrow P^{ss}(x)&=&\frac{1}{\sqrt{2\,\pi\,W_{\Omega}}}
e^{-\frac{x^2}{2\,W_{\Omega}}}.
\end{eqnarray}

Similarly, it is straightforward to show that:
\begin{eqnarray}\label{eq.32}
P^{ss}(v)&=&\frac{1}{\sqrt{2\,\pi\,W_{\Theta}}}
e^{-\frac{v^2}{2\,W_{\Theta}}}.
\end{eqnarray}

\subsection{Stationary distribution}

The Eqs.(\ref{eq.29}) and~(\ref{eq.32}) combine to
\begin{equation}
P^{ss}(x,v)=P^{ss}(x)P^{ss}(v)=\frac{1}{2\,\pi\,\sqrt{W_{\Theta}\,W_{\Omega}}}
\exp\left\{-\frac12\left(\frac{x^2}{W_{\Omega}} +
\frac{v^2}{W_{\Theta}} \right)\right\}.\label{Pss}
\end{equation}

The distribution on Eq.(\ref{Pss}) is complete, normalized and
exact. It represents the stationary state for extremely long times
and includes more information than similar models in the massless
(over\,--\,damped) limit. Since $W_{\Omega}\neq W_{\Theta}$, a
renormalization of the mass and of the temperature will occur.
Recent works have suggested other mechanisms for deviations from
Boltzmann statistics~\cite{2006_EL_75_15, 2007_EL_79_60003,
2007_PRE_76_030101}. A massless BP is bound to move as the
instantaneous force directs it, but an inertial one can accumulate
kinetic energy and its velocity deviates from the ratio between
the external force and the friction coefficient, creating an
opportunity for a feed-back mechanism (incipiently present in the
non-Markovian colored noise $\xi$ time-scale $\tau$) to modify the
stationary distribution away from the Boltzmann equilibrium
form~\cite{2006_PhysA_365_289, 1999_PRE_60_5208, 1982_PRA_26_1589,
1989_PRA_40_7312} even at $T_1=T_2$~\cite{2006_PhysA_365_289,
1999_PRE_60_5208}. Taking the limit $m\rightarrow 0$ at $T_1=T_2$
restores the Boltzmann equilibrium form, despite $\tau>0$, due to
the crucial role of inertia in the mechanism above.

\section{Over\,--\,damped limit}

The Eq.(\ref{eq.28}) defining the $W_{\Omega}$ term, (see
Appendix~\ref{appb} for a complete derivation), gives the
dependence of the stationary position distribution on all
parameters of the system in its most general form. This
distribution is a generalization of the expression obtained in
Ref.~\cite{2006_PhysA_365_289} and, similarly to that case, it
differs from the Boltzmann form due to the non-Markovian character
of the colored noise, but tends to it when we take
$\tau\rightarrow0$ and $T_1=T_2$. The nontrivial forms of
Eqs.(\ref{eq.28}) and (\ref{eq.31}) indicate that a
renormalization of the mass (or the rigidity $k$) takes place due
to the non-Markovian character of $\xi$ ~\cite{1982_PRA_26_1589,
1989_PRA_40_7312, 2006_PhysA_365_289}.

In the over damped limit, $m\rightarrow 0$, the term $W_{\Omega}$
simplifies considerably and reads
\begin{equation}\label{eq.34}
\lim_{m\rightarrow0}W_{\Omega}=\,{\frac
{{T_1}\,\tau\,k+{\Gamma_1}\,{T_1}+{\Gamma_2}\,{ T_2}}{ \left(
{\Gamma_1}+{\Gamma_2}+k\tau \right) k}},
\end{equation}

The quantity $k\,W_{\Omega}$ behaves as the exact effective
temperature $T_{eff}$ for the system in the over\,--\,damped
limit. We observe that both the slow and the fast noise contribute
to $T_{eff}$. It can be seen that $T_{eff} = k\,W_{\Omega}$ has an
intermediate value between $T_1$ and $T_2$:
\[
\mbox{min}[T_1,T_2]\leq k\,W_{\Omega}\leq \mbox{max}[T_1,T_2].
\]

Comparing Eq.(\ref{eq.34}) with other approaches, we noticed that
in Ref.~\cite{2000_JPSJ_69_247} the authors show the steps to
obtain the stationary distribution for a similar model in the over
damped limit. In their case, they make approximations in such a
way that both the slow field $\xi$ and the dissipative memory
term, proportional to $\Gamma_2$ ($\Gamma_1$ in their notation),
behave as if they were effective external additive terms for the
potential $V(x)$. In consequence, the frictional coefficient
$\Gamma_1$ ($\Gamma_0$ in their notation) will not contribute to
the equilibrium distribution since that is a function of the
effective external terms only. The system is supposed to quickly
adjust to that potential, giving rise to an instantaneous
equilibrium distribution.

On the other hand, in the present method we calculate the averages
rather differently. As we have seen, we first integrate over the
noise-average of the exact distribution, Eq.(\ref{eq.08}), for an
infinitely long-time and then obtain the time-average of them. By
doing this, the neglected effect of $\Gamma_1$ becomes apparent
and shows in our solution. So, in order to recover the results
suggested on section 4.1 on Ref.~\cite{2000_JPSJ_69_247} (derived
in appendix~\ref{appc}) all we need to do is take $\Gamma_1=0$ in
Eq.(\ref{eq.34}). In our notation, the results on
Ref.~\cite{2000_JPSJ_69_247} read (where we assume
$V(x)=\frac{k\,x^2}{2}$ so our models coincide)
\[
W_{\Omega}^{[15]}=\,{\frac {{T_1}\,\tau\,k+{\Gamma_2}\,{T_2}}{
\left( {\Gamma_2}+k\tau \right) k}},
\]
which is identical to ours when we take $\Gamma_1=0$ in
Eq.(\ref{eq.34}).

When $T_1=T_2=T$ we recover the Boltzmann equilibrium with
distribution
\begin{equation}\label{eq.35}
W_{\Omega}= \frac{T}{k}\Rightarrow
P^{ss}(x)=\sqrt{\frac{k}{2\,\pi\,T}} e^{-\frac{kx^2}{2\,T}}.
\end{equation}

For the velocity distribution, in the over\,--\,damped limit it
assumes a simple form. As is the case of $W_{\Omega}$, the
expression for $W_{\Theta}$ in Eq.(\ref{eq.31}) expresses the
dependence of the velocity distribution on all parameters of the
model. However, by taking $m\rightarrow0$ in Eq.(\ref{eq.31})
gives asymptotically:
\begin{equation}
\lim_{m\rightarrow0}\,W_{\Theta}\,\rightarrow\,\frac{T_1}{m}.
\end{equation}
We notice that only the fast noise $T_1$ contributes to the
velocity distribution. The kinetic energy of the BP is completely
driven by the fast noise and ignores the slow one in the
over\,--\,damped limit.

The normalized velocity distribution reads in that approximation
order:
\begin{equation}\label{eq.36}
W_{\Theta}=\frac{T_1}{m}\Rightarrow
P^{ss}(v)=\sqrt{\frac{m}{2\,\pi\,T_1}} e^{-\frac{mv^2}{2\,T_1}}.
\end{equation}
In this limit we observe that only the fast noise
($T_1,\,\Gamma_1$) drives the kinetic energy distribution.

\section{Conclusions}
A better understanding of slow dynamics and associated models may
require powerful numerical methods, and computers, as one tries to
simulate the (extremely) long-time relaxation that often occurs
for interesting systems such as glasses.

In this context, simpler models may be very useful for obtaining
some good qualitative understanding of the long time limit without
presenting the numerical difficulties the more realistic ones do,
with the possibility of obtaining exact results. These are always
interesting in that they take into account all the physical
effects present in a given model, independently of any
approximation. In other words, exact treatments contain all the
available information about a model. Some of it being inaccessible
through approximate methods.

In that spirit, BP models have been proposed as a means to study
some of the phenomena associated with the competition between fast
thermal fluctuations and slow structural relaxation in
glasses~\cite{2000_EPJB_16_317, 2000_JPSJ_69_247, kurchan_nature}.
A white noise (fast) is associated with the thermal fluctuations
that happen in short time-scales. A colored noise function (slow)
is associated with the long time relaxation (structural).

For these simple models, a typical approach is to consider the
over-damped regime, which is equivalent to the case when the BP
has zero mass ($m\rightarrow 0^{+}$). When responding to external
forces acting on it the BP's velocity assumes the ratio between
the sum of all external forces and a friction coefficient. These
models can be used to study the emergence of a non-equilibrium
stationary state as the probability distribution tends to the
stationary form as $t\rightarrow \infty$.

In the present work, we develop an exact method for time-averaging
the distribution for $x$ and $v$ for a BP submitted to a fast
white noise and a slow non-Markovian one at different
temperatures.

The essence of our method is the use of time-averages of the
distributions for $x$ and $v$ which leads us naturally to the use
of Laplace transformations in order to ``open'' the problem. In a
straightforward way, we represent the time-averaged distribution
as a sum of integrals that can be easily analyzed, and computed
exactly. The distribution thus obtained can be compared to the
ones in the literature. By being exact, our method allows us to
obtain information about the effects due to the finite mass of the
BP showing, for instance, that the fast noise term plays a role on
the stationary distribution of positions that is usually neglected
in other methods.

\begin{acknowledgments}
The authors are grateful to C. Anteneodo and J.L. Gonzalez for
reading and commenting the manuscript. One of us W.A.M.M. thanks
the Brazilian funding agencies Faperj and CNPq and D.O.S.P. would
like to thank the Brazilian funding agency CAPES for the financial
support at Universidade de Aveiro at Portugal.
\end{acknowledgments}

\newpage

\appendix
\setcounter{equation}{0}
\section{}\label{appa}
The Laplace transformation of Eq.(\ref{eq.26}) is shown
below~\cite{2006_PhysA_365_289}. We will give a step by step
derivation for the expression of $P^{ss}(x,v)$ as a function of
the Laplace transforms of the position and velocity for the
Brownian particle.

Lets start with the basic definition:
\begin{eqnarray*}
P^{ss}(x,v)&=&\lim_{z\rightarrow0^{+}}z
\int_{0}^{\infty}e^{-zt}\langle{\delta}(x-x(t)){\delta}(v-v(t))\rangle
dt.
\end{eqnarray*}

We write the delta-functions above in the integral representation:
\begin{eqnarray*}
\langle\delta(x-x(t))\delta(v-v(t))\rangle &=&
\int_{-\infty}^{+\infty}\frac{dQ}{2\pi}e^{iQx}
\int_{-\infty}^{+\infty}\frac{dP}{2\pi}e^{iPv}
\sum_{l=0}^{\infty}\frac{(-iQ)^{l}}{l!}
\sum_{m=0}^{\infty}\frac{(-iP)^{m}}{m!} \langle
x^{l}(t)v^{m}(t)\rangle,
\end{eqnarray*}
and obtain (after using the delta functions to express identically
the averages over the noise as functions at distinct times):
\begin{eqnarray*}
P^{ss}(x,v) &=&\lim_{z\rightarrow0^{+}}z
\int_{0}^{\infty}dte^{-zt}\int_{-\infty}^{+\infty}
\frac{dQ}{2\pi}e^{iQx}\int_{-\infty}^{+\infty}\frac{dP}{2\pi}e^{iPv}
\sum_{l=0}^{\infty}\frac{(-iQ)^{l}}{l!}\sum_{m=0}^{\infty}\frac{(-iP)^{m}}{m!}
\nonumber\\ &\times &
\int_{0}^{\infty}\prod_{f=1}^{l}dt_{lf}\delta(t-t_{lf})
\int_{0}^{\infty}\prod_{h=1}^{m}dt_{mh}\delta(t-t_{mb}) \langle
\prod_{f=1}^{l} x(t_{lf}) \prod_{h=1}^{m}v(t_{mh})\rangle.
\end{eqnarray*}

Next, we express all delta functions above as integrals on the
complex plane, displaced from the complex axis by a factor of
$\epsilon$ (that vanishes faster than $z$). That factor will
guarantee the convergence of the Laplace transforms for positions
and velocities in the following.
\begin{eqnarray*}
P^{ss}(x,v)
&=&\lim_{z,\epsilon\rightarrow0^{+}}z \int_{0}^{\infty}dte^{-zt}
\int_{-\infty}^{+\infty}\frac{dQ}{2\pi}e^{iQx}
\int_{-\infty}^{+\infty}\frac{dP}{2\pi}e^{iPv}
\sum_{l=0}^{\infty}\frac{(-iQ)^{l}}{l!}\sum_{m=0}^{\infty}
\frac{(-iP)^{m}}{m!} \nonumber\\ &\times
&\int_{-\infty}^{+\infty}\prod_{f=1}^{l}\frac{dq_{f}}{2\pi}
\prod_{h=1}^{m}\frac{dp_{h}}{2\pi}
\int_{0}^{\infty}\prod_{f=1}^{l}dt_{lf}
\int_{0}^{\infty}\prod_{h=1}^{m}dt_{mh} \nonumber\\ &\times&
e^{\sum_{f=1}^{l}(t-t_{lf})(iq_{f}+\epsilon)
+\sum_{h=1}^{m}(t-t_{mh})(ip_{h}+\epsilon)} \langle
\prod_{f=1}^{l} x(t_{lf}) \prod_{h=1}^{m}v(t_{mh})\rangle.
\end{eqnarray*}

We need to integrate over all $\left\{t_{lf},t_{mh}\right\}$,
obtaining the averages over the Laplace transforms of the position
and velocity:
\begin{eqnarray*}
P^{ss}(x,v)
&=&\lim_{z,\epsilon\rightarrow0^{+}}\int_{-\infty}^{+\infty}
\frac{dQ}{2\pi}e^{iQx} \int_{-\infty}^{+\infty}
\frac{dP}{2\pi}e^{iPv} \sum_{l=0}^{\infty}
\frac{(-iQ)^{l}}{l!}\sum_{m=0}^{\infty} \frac{(-iP)^{m}}{m!}
\int_{-\infty}^{+\infty} \prod_{f=1}^{l}\frac{dq_{f}}{2\pi}
\prod_{h=1}^{m} \frac{dp_{h}}{2\pi}
 \nonumber\\ &\times&
\int_{0}^{\infty}dt\,z\,
e^{-t\left\{z-\sum_{f=1}^{l}(iq_{f}+\epsilon)
-\sum_{h=1}^{m}(ip_{h}+\epsilon)\right\}}
\langle\prod_{f=1}^{l}\tilde{x}(iq_{f}+\epsilon)\prod_{h=1}^{m}
\tilde{v}(ip_{h}+\epsilon)\rangle.
\end{eqnarray*}

Finally, we integrate over $t$ and obtain Eq.(\ref{eq.26}):
\begin{eqnarray}\label{eq.b1}
P^{ss}(x,v) &=&\lim_{z,\epsilon\rightarrow0^{+}}
\sum_{l,m=0}^{\infty} \int_{-\infty}^{+\infty}\frac{dQ}{2\pi}
\frac{dP}{2\pi}e^{iQx+iPv} \frac{(-iQ)^{l}}{l!}
\frac{(-iP)^{m}}{m!} \int_{-\infty}^{+\infty}
\prod_{f=1}^{l}\frac{dq_{f}}{2\pi}
\prod_{h=1}^{m}\frac{dp_{h}}{2\pi} \nonumber\\
&\times&\frac{z}{z-\left[\sum_{f=1}^{l}iq_{f}+
\sum_{h=1}^{m}ip_{h}+(l+m)\epsilon\right]}
\langle\prod_{f=1}^{l}\tilde{x}(iq_{f}+\epsilon)\prod_{h=1}^{m}
\tilde{v}(ip_{h}+\epsilon)\rangle
\end{eqnarray}

\newpage

\section{}\label{appb}

A typical calculation of the stationary state distribution of
displacements terms is shown below for the $W_{\Omega}$ term from
Eq.(\ref{eq.27}).

The first integral in Eq.(\ref{eq.27}) reads:
\begin{eqnarray*}
&& \int_{-\infty}^{+\infty}\frac{dq_{i}}{2\pi}\frac{dq_{j}}{2\pi}
\frac{z}{z-i(q_{i}+q_{j}-2i\,\epsilon+\diamond)}\,\Omega(iq_{i}+\epsilon)\Omega(iq_{j}+\epsilon)
\langle\tilde{\eta}(iq_{i}+\epsilon)\tilde{\eta}(iq_{j}+\epsilon)\rangle
=\nonumber\\
&=& \int_{-\infty}^{+\infty}\frac{dq_{i}}{2\pi}\frac{dq_{j}}{2\pi}
\frac{z}{z-i(q_{i}+q_{j}-2i\,\epsilon+\diamond)}\,\frac{1+\tau\,(iq_{i}+\epsilon)}
{[q_{i}-i(\epsilon-z_{1})]\,[q_{i}-i(\epsilon-z_{2})]\,[q_{i}-i(\epsilon-z_{3})]}
\nonumber\\ &\times& \frac{1+\tau\,(iq_{i}+\epsilon)}
{[q_{j}-i(\epsilon-z_{1})]\,[q_{j}-i(\epsilon-z_{2})]\,[q_{j}-i(\epsilon-z_{3})]}\,
\frac{\Gamma_{1}\,T_{1}\,m^{-2}\,\tau^{-2}}{(-i)[q_{j}-(-q_{i}+2\,i\,\epsilon)]}
\end{eqnarray*}
To continue the calculation, the integrations over the poles must
be done in a way that allows us to obtain, after all integrations
have been done (as explained in Section III.A),
$\lim_{z\rightarrow 0} I(z)=1$. When integrating over the poles of
$q_{j}$'s it is possible to see, Fig.(\ref{fig}), that it will
work only for $q_{j}=-q_{i}+2\,i\,\epsilon$. Thus:
\begin{eqnarray*}
&\Rightarrow&
\int_{-\infty}^{+\infty}\frac{dq_{i}}{2\pi}\frac{dq_{j}}{2\pi}
\frac{z}{z-i(q_{i}+q_{j}-2i\,\epsilon+\diamond)}\,\Omega(iq_{i}+\epsilon)\Omega(iq_{j}+\epsilon)
\langle\tilde{\eta}(iq_{i}+\epsilon)\tilde{\eta}(iq_{j}+\epsilon)\rangle
=\nonumber\\
&=&\frac{\Gamma_{1}\,T_{1}}{m^{2}\,\tau^{2}}\,\frac{z}{z-i\diamond}\,
\int_{-\infty}^{+\infty}\frac{dq_{i}}{2\pi}
\frac{1-\tau^{2}\,(iq_{i}+\epsilon)^{2}}
{[q_{i}-i(\epsilon-z_{1})]\,[q_{i}-i(\epsilon-z_{2})]\,[q_{i}-i(\epsilon-z_{3})]\,
[q_{i}-i(\epsilon+z_{1})]\,[q_{i}-i(\epsilon+z_{2})]\,[q_{i}-i(\epsilon+z_{3})]}
\end{eqnarray*}

The same holds for the integration over the $q_{i}$'s poles. So,
the non-zero contribution comes only from the poles $q_{i} =
i(\epsilon-z_{\alpha}), \alpha = 1,2,3$, in the upper part of
Fig.(\ref{fig}):
\begin{eqnarray*}
&& \int_{-\infty}^{+\infty}\frac{dq_{i}}{2\pi}\frac{dq_{j}}{2\pi}
\frac{z}{z-i(q_{i}+q_{j}-2i\,\epsilon+\diamond)}\,\Omega(iq_{i}+\epsilon)\Omega(iq_{j}+\epsilon)
\langle\tilde{\eta}(iq_{i}+\epsilon)\tilde{\eta}(iq_{j}+\epsilon)\rangle
=\nonumber\\
&=&\frac{\Gamma_{1}\,T_{1}}{m^{2}\,\tau^{2}}\,\frac{z}{z-i\diamond}\,\left\{
\frac{\tau^{2}\,z_{1}^{2}-1}{z_{1}\,\mid z_{1}-z_{2}\mid^{2}\,\mid
z_{1}+z_{2}\mid^{2}}-\frac{\tau^{2}[z_{2}\,(z_{1}-z_{2})^{*}(z_{1}+z_{2})^{*}-z_{2}^{*}\,(z_{1}-z_{2})(z_{1}+z_{2})]}
{4\,i\,\Re(z_{2})\,\Im(z_{2})\mid z_{1}-z_{2}\mid^{2}\,\mid
z_{1}+z_{2}\mid^{2}}\right. \nonumber\\
&+&\left.\frac{[z_{2}^{*}(z_{1}-z_{2})^{*}(z_{1}+z_{2})^{*}-z_{2}(z_{1}-z_{2})(z_{1}+z_{2})]}
{4\,i\,\Re(z_{2})\,\Im(z_{2})\mid z_{2}\mid^{2}\mid
z_{1}-z_{2}\mid^{2}\,\mid z_{1}+z_{2}\mid^{2}}\right\}
\end{eqnarray*}

Finally, using the definitions of Eq.(\ref{eq.25}), we get for the
first part of Eq.(\ref{eq.27}):
\begin{eqnarray*}
&& \int_{-\infty}^{+\infty}\frac{dq_{i}}{2\pi}\frac{dq_{j}}{2\pi}
\frac{z}{z-i(q_{i}+q_{j}-2i\,\epsilon+\diamond)}\,\Omega(iq_{i}+\epsilon)\Omega(iq_{j}+\epsilon)
\langle\tilde{\eta}(iq_{i}+\epsilon)\tilde{\eta}(iq_{j}+\epsilon)\rangle
=\nonumber\\
&=&\frac{z}{z-i\diamond}\,\left\{\frac{\Gamma_{1}\,T_{1}}{2\,m^{2}}\,
\frac{2\,z_{1}\,\mathcal{A\,B}+(\mathcal{B\,D\,G-B\,C\,F})+(\mathcal{A\,D\,F+A\,C\,G})}
{\mathcal{A\,B}\,(\mathcal{C}^{2}+\mathcal{D}^{2})\,(\mathcal{F}^{2}+\mathcal{G}^{2})}\right.
\nonumber\\
&-&\left. \frac{\Gamma_{1}\,T_{1}}{2\,m^{2}\,\tau^{2}}\,
\frac{2\,\mathcal{A\,B}\,(\mathcal{A}^{2}+\mathcal{B}^{2})-z_{1}\,(\mathcal{B\,D\,G-B\,C\,F})
+z_{1}\,(\mathcal{A\,D\,F+A\,C\,G})}
{z_{1}\,\mathcal{A\,B}\,(\mathcal{A}^{2}+\mathcal{B}^{2})\,(\mathcal{C}^{2}+\mathcal{D}^{2})
\,(\mathcal{F}^{2}+\mathcal{G}^{2})}\right\}
\end{eqnarray*}

For the second part of Eq.(\ref{eq.27}), the second term of the
RHS of Eq.(\ref{eq.13}) does not contribute since it is
straightforward to show that it leads to a null contribution.
Thus, we shall compute the contribution from the first part of
Eq.(\ref{eq.27}) keeping only the RHS of Eq.(\ref{eq.13}). It
yields:
\begin{eqnarray}\label{eq.cc1}
&& \int_{-\infty}^{+\infty}\frac{dq_{i}}{2\pi}\frac{dq_{j}}{2\pi}
\frac{z}{z-i(q_{i}+q_{j}-2i\,\epsilon+\diamond)}\,\Omega(iq_{i}+\epsilon)\Omega(iq_{j}+\epsilon)
\langle\tilde{\xi}(iq_{i}+\epsilon)\tilde{\xi}(iq_{j}+\epsilon)\rangle
= \nonumber\\
&=& \int_{-\infty}^{+\infty}\frac{dq_{i}}{2\pi}\frac{dq_{j}}{2\pi}
\frac{z}{z-i(q_{i}+q_{j}-2i\,\epsilon+\diamond)}\,\frac{1+\tau\,(iq_{i}+\epsilon)}
{[q_{i}-i(\epsilon-z_{1})]\,[q_{i}-i(\epsilon-z_{2})]\,[q_{i}-i(\epsilon-z_{3})]}
\nonumber\\ &\times& \frac{1+\tau\,(iq_{i}+\epsilon)}
{[q_{j}-i(\epsilon-z_{1})]\,[q_{j}-i(\epsilon-z_{2})]\,[q_{j}-i(\epsilon-z_{3})]}\,
\frac{\Gamma_{2}\,T_{2}\,m^{-2}\,\tau^{-2}}{(-i)[q_{j}-(-q_{i}+2\,i\,\epsilon)][1-\tau(iq_{i}+\epsilon)]
[1-\tau(iq_{j}+\epsilon)]}
\end{eqnarray}
The only contributing integrations will be the ones over the poles
$q_{j}=-q_{i}+2\,i\,\epsilon$, (see Fig.(\ref{fig})). So:

\begin{eqnarray*}
&\Rightarrow&
\int_{-\infty}^{+\infty}\frac{dq_{i}}{2\pi}\frac{dq_{j}}{2\pi}
\frac{z}{z-i(q_{i}+q_{j}-2i\,\epsilon+\diamond)}\,\Omega(iq_{i}+\epsilon)\Omega(iq_{j}+\epsilon)
\langle\tilde{\xi}(iq_{i}+\epsilon)\tilde{\xi}(iq_{j}+\epsilon)\rangle
= \nonumber\\
&=&
\frac{\Gamma_{2}\,T_{2}}{m^{2}\,\tau^{2}}\,\frac{z}{z-i\diamond}\,
\int_{-\infty}^{+\infty}\frac{dq_{i}}{2\pi} \frac{1}
{[q_{i}-i(\epsilon-z_{1})]\,[q_{i}-i(\epsilon-z_{2})]\,[q_{i}-i(\epsilon-z_{3})]}
\frac{1}{[q_{i}-i(\epsilon+z_{1})]\,[q_{i}+i(\epsilon-z_{2})]\,[q_{i}+i(\epsilon-z_{3})]}
\end{eqnarray*}
Again, the contribution came from the poles $q_{i} =
i(\epsilon-z_{\alpha}), \alpha = 1,2,3$. So:
\begin{eqnarray*}
&& \int_{-\infty}^{+\infty}\frac{dq_{i}}{2\pi}\frac{dq_{j}}{2\pi}
\frac{z}{z-i(q_{i}+q_{j}-2i\,\epsilon+\diamond)}\,\Omega(iq_{i}+\epsilon)\Omega(iq_{j}+\epsilon)
\langle\tilde{\xi}(iq_{i}+\epsilon)\tilde{\xi}(iq_{j}+\epsilon)\rangle
= \nonumber\\
&=&
\frac{\Gamma_{2}\,T_{2}}{2\,m^{2}\,\tau^{2}}\,\frac{z}{z-i\diamond}\,
\left\{\frac{-1} {z_{1}\,\mid z_{1}-z_{2}\mid^{2}\,\mid
z_{1}+z_{2}\mid^{2}}+\frac{[z_{2}^{*}(z_{1}-z_{2})^{*}(z_{1}+z_{2})^{*}-z_{2}(z_{1}-z_{2})(z_{1}+z_{2})]}
{4\,i\,\Re(z_{2})\,\Im(z_{2})\mid z_{2}\mid^{2}\mid
z_{1}-z_{2}\mid^{2}\,\mid z_{1}+z_{2}\mid^{2}}\right\}.
\end{eqnarray*}

Using the definitions on Eq.(\ref{eq.25}) we obtain:
\begin{eqnarray}\label{eq.cc2}
&&\int_{-\infty}^{+\infty}\frac{dq_{i}}{2\pi}\frac{dq_{j}}{2\pi}
\frac{z}{z-i(q_{i}+q_{j}-2i\,\epsilon+\diamond)}\,\Omega(iq_{i}+\epsilon)\Omega(iq_{j}+\epsilon)
\langle\tilde{\xi}(iq_{i}+\epsilon)\tilde{\xi}(iq_{j}+\epsilon)\rangle
= \nonumber\\
&=& -\frac{\Gamma_{2}\,T_{2}}{2\,m^{2}\,\tau^{2}}\,
\frac{2\,\mathcal{A\,B}\,(\mathcal{A}^{2}+\mathcal{B}^{2})-z_{1}\,(\mathcal{B\,D\,G-B\,C\,F})
+z_{1}\,(\mathcal{A\,D\,F+A\,C\,G})}
{z_{1}\,\mathcal{A\,B}\,(\mathcal{A}^{2}+\mathcal{B}^{2})\,(\mathcal{C}^{2}+\mathcal{D}^{2})
\,(\mathcal{F}^{2}+\mathcal{G}^{2})}
\end{eqnarray}

Combining Eqs.(\ref{eq.cc1}) and (\ref{eq.cc2}) we obtain the
result for $W_{\Omega}$ in Eq.(\ref{eq.28}). An analogous
calculation can be done for the velocity integrals and obtain
$W_{\Theta}$, e.g., Eq.(\ref{eq.31}).

\newpage

\section{}\label{appc}

The stationary distribution of $x$ is obtained in Cugliandolo and
Kurchan~\cite{2000_JPSJ_69_247}. We review their main steps below.
Let's define:
\begin{equation}\label{eq.c1}
P(x) = \int dh P(x/h)P(h),
\end{equation}
where
\begin{equation}\label{eq.c2}
P(h) =
\frac{e^{-\beta^{*}(F(h)+\frac{h^{2}}{2\,\frac{\Gamma_{2}}{\tau}})}}{\int
dh\,e^{-\beta^{*}(F(h)+\frac{h^{2}}{2\,\frac{\Gamma_{2}}{\tau}})}},
\end{equation}
and
\begin{equation}\label{eq.c3}
P(x/h)=\frac{e^{-\beta(V(x)+\frac{\Gamma_{2}}{\tau}\,\frac{x^2}{2}-h\,x)}}{\int
dx\,e^{-\beta(V(x)+\frac{\Gamma_{2}}{\tau}\,\frac{x^2}{2}-h\,x)}}.
\end{equation}
The denominator above defines $Z(h)$ and $F(h) \equiv
-\beta^{-1}\ln Z(h)$. For $V(x) = k\frac{x^2}{2}$, we have:
\begin{equation}\label{eq.c4}
Z(h) = \sqrt{\frac{2\,\pi}{\beta\, (k+\frac{\Gamma_{2}}{\tau})}}\,
e^{\frac{\beta\,h^2}{2\,(k+\frac{\Gamma_{2}}{\tau})}}; \,\, F(h) =
-\frac{h^2}{2\,(k+\frac{\Gamma_{2}}{\tau})}-\beta^{-1}C
\end{equation}
And the denominator of Eq.(\ref{eq.c2}) becomes:
\begin{equation}\label{eq.c6}
\int
dh\,e^{-\beta^{*}\left(-\frac{h^2}{2\,(k+\frac{\Gamma_{2}}{\tau})}
-\beta^{-1}C+\frac{h^{2}}{2\,\frac{\Gamma_{2}}{\tau}}\right)}=
e^{\frac{\beta^{*}\,C}{\beta}}
\sqrt{\frac{2\,\pi\,\frac{\Gamma_{2}}{\tau}\,
(k+\frac{\Gamma_{2}}{\tau})}{\beta^{*}\,k}}
\end{equation}
Thus, Eq.(\ref{eq.c1}) becomes:
\begin{equation}\label{eq.c8}
P(x) = \sqrt{\frac{\beta^{*}\,\beta\,\,k\,
(k\tau+\Gamma_{2})}{2\,\pi\,
(\beta^{*}\,k\tau+\beta\,\Gamma_{2})}} \,
e^{-\frac{\beta^{*}\,\beta\,\,
k\,(k\tau+\Gamma_{2})\,x^2}{2\,(\beta^{*}k\tau+\beta\,\Gamma_{2})}}
\end{equation}

When the temperatures are equal, then $\beta^{*}=\beta = 1/T$ and:
\begin{equation}\label{eq.c9}
P(x)= \sqrt{\frac{k}{2\,\pi\,T}} \,e^{-\frac{k\,x^2}{2\,T}}
\end{equation}

\newpage

\newpage


\begin{thebibliography}{32}
\expandafter\ifx\csname
natexlab\endcsname\relax\def\natexlab#1{#1}\fi
\expandafter\ifx\csname bibnamefont\endcsname\relax
  \def\bibnamefont#1{#1}\fi
\expandafter\ifx\csname bibfnamefont\endcsname\relax
  \def\bibfnamefont#1{#1}\fi
\expandafter\ifx\csname citenamefont\endcsname\relax
  \def\citenamefont#1{#1}\fi
\expandafter\ifx\csname url\endcsname\relax
  \def\url#1{\texttt{#1}}\fi
\expandafter\ifx\csname
urlprefix\endcsname\relax\def\urlprefix{URL }\fi
\providecommand{\bibinfo}[2]{#2}
\providecommand{\eprint}[2][]{\url{#2}}

\bibitem[{\citenamefont{Das}(2004)}]{RMP_76_2004_785}
\bibinfo{author}{\bibfnamefont{S.~P.} \bibnamefont{Das}},
  \bibinfo{journal}{Rev. Mod. Phys.} \textbf{\bibinfo{volume}{76}},
  \bibinfo{pages}{785} (\bibinfo{year}{2004}).

\bibitem[{\citenamefont{{van Kampen} and
  {Oppenheim}}(1986)}]{1986_PhysA_138_231}
\bibinfo{author}{\bibfnamefont{N.~G.} \bibnamefont{{van Kampen}}}
  \bibnamefont{and}
  \bibinfo{author}{\bibfnamefont{I.}~\bibnamefont{{Oppenheim}}},
  \bibinfo{journal}{Physica A} \textbf{\bibinfo{volume}{138}},
  \bibinfo{pages}{231} (\bibinfo{year}{1986}).

\bibitem[{\citenamefont{{Cugliandolo} and {Kurchan}}(2000)}]{2000_JPSJ_69_247}
\bibinfo{author}{\bibfnamefont{L.}~\bibnamefont{{Cugliandolo}}}
  \bibnamefont{and}
  \bibinfo{author}{\bibfnamefont{J.}~\bibnamefont{{Kurchan}}},
  \bibinfo{journal}{Journal of the Physical Society of Japan}
  \textbf{\bibinfo{volume}{69}}, \bibinfo{pages}{247} (\bibinfo{year}{2000}),
  \eprint{arXiv:cond-mat/9911086}.

\bibitem[{\citenamefont{{Allahverdyan} and
  {Nieuwenhuizen}}(2000)}]{2000_PRE_62_845}
\bibinfo{author}{\bibfnamefont{A.~E.} \bibnamefont{{Allahverdyan}}}
  \bibnamefont{and} \bibinfo{author}{\bibfnamefont{T.~M.}
  \bibnamefont{{Nieuwenhuizen}}}, \bibinfo{journal}{Physical Review E}
  \textbf{\bibinfo{volume}{62}}, \bibinfo{pages}{845} (\bibinfo{year}{2000}).

\bibitem[{\citenamefont{{Frank}}(2007)}]{2007_PLA_360_552}
\bibinfo{author}{\bibfnamefont{T.~D.} \bibnamefont{{Frank}}},
  \bibinfo{journal}{Physics Letters A} \textbf{\bibinfo{volume}{360}},
  \bibinfo{pages}{552} (\bibinfo{year}{2007}).

\bibitem[{\citenamefont{{Frank}}(2006)}]{2006_PLA_357_275}
\bibinfo{author}{\bibfnamefont{T.~D.} \bibnamefont{{Frank}}},
  \bibinfo{journal}{Physics Letters A} \textbf{\bibinfo{volume}{357}},
  \bibinfo{pages}{275} (\bibinfo{year}{2006}).

\bibitem[{\citenamefont{{Frank}}(2005)}]{2005_PRE_72_011112}
\bibinfo{author}{\bibfnamefont{T.~D.} \bibnamefont{{Frank}}},
  \bibinfo{journal}{Physical Review E} \textbf{\bibinfo{volume}{72}},
  \bibinfo{pages}{011112} (\bibinfo{year}{2005}).

\bibitem[{\citenamefont{{Ritter} et~al.}(2004)\citenamefont{{Ritter},
  {D'Ajello}, and {Figueiredo}}}]{2004_PRE_69_016119}
\bibinfo{author}{\bibfnamefont{O.~M.} \bibnamefont{{Ritter}}},
  \bibinfo{author}{\bibfnamefont{P.~C.} \bibnamefont{{D'Ajello}}},
  \bibnamefont{and}
  \bibinfo{author}{\bibfnamefont{W.}~\bibnamefont{{Figueiredo}}},
  \bibinfo{journal}{Physical Review E} \textbf{\bibinfo{volume}{69}},
  \bibinfo{pages}{016119} (\bibinfo{year}{2004}).

\bibitem[{\citenamefont{{Zamponi} et~al.}(2005)\citenamefont{{Zamponi},
  {Bonetto}, {Cugliandolo}, and {Kurchan}}}]{2005_JSTAT_09_P09013}
\bibinfo{author}{\bibfnamefont{F.}~\bibnamefont{{Zamponi}}},
  \bibinfo{author}{\bibfnamefont{F.}~\bibnamefont{{Bonetto}}},
  \bibinfo{author}{\bibfnamefont{L.~F.} \bibnamefont{{Cugliandolo}}},
  \bibnamefont{and}
  \bibinfo{author}{\bibfnamefont{J.}~\bibnamefont{{Kurchan}}},
  \bibinfo{journal}{Journal of Statistical Mechanics: Theory and Experiment}
  \textbf{\bibinfo{volume}{9}}, \bibinfo{pages}{13} (\bibinfo{year}{2005}),
  \eprint{arXiv:cond-mat/0504750}.

\bibitem[{\citenamefont{{Cugliandolo} and
  {Kurchan}}(1999)}]{1999_PhysA_263_242}
\bibinfo{author}{\bibfnamefont{L.}~\bibnamefont{{Cugliandolo}}}
  \bibnamefont{and}
  \bibinfo{author}{\bibfnamefont{J.}~\bibnamefont{{Kurchan}}},
  \bibinfo{journal}{Physica A} \textbf{\bibinfo{volume}{263}},
  \bibinfo{pages}{242} (\bibinfo{year}{1999}), \eprint{arXiv:cond-mat/9807226}.

\bibitem[{\citenamefont{{Ilg} and {Barrat}}(2006)}]{2006_JPhysCS_40_76}
\bibinfo{author}{\bibfnamefont{P.}~\bibnamefont{{Ilg}}} \bibnamefont{and}
  \bibinfo{author}{\bibfnamefont{J.-L.} \bibnamefont{{Barrat}}},
  \bibinfo{journal}{Journal of Physics Conference Series}
  \textbf{\bibinfo{volume}{40}}, \bibinfo{pages}{76} (\bibinfo{year}{2006}),
  \eprint{arXiv:cond-mat/0601618}.

\bibitem[{\citenamefont{{Budini} and {C{\'a}ceres}}(2004)}]{2004_PRE_70_046104}
\bibinfo{author}{\bibfnamefont{A.~A.} \bibnamefont{{Budini}}} \bibnamefont{and}
  \bibinfo{author}{\bibfnamefont{M.~O.} \bibnamefont{{C{\'a}ceres}}},
  \bibinfo{journal}{Physical Review E} \textbf{\bibinfo{volume}{70}},
  \bibinfo{pages}{046104} (\bibinfo{year}{2004}).

\bibitem[{\citenamefont{{C{\'a}ceres}}(2003)}]{2003_PRE_67_016102}
\bibinfo{author}{\bibfnamefont{M.~O.} \bibnamefont{{C{\'a}ceres}}},
  \bibinfo{journal}{Physical Review E} \textbf{\bibinfo{volume}{67}},
  \bibinfo{pages}{016102} (\bibinfo{year}{2003}).

\bibitem[{\citenamefont{{Sandri}}(1963{\natexlab{a}})}]{sandri1}
\bibinfo{author}{\bibfnamefont{G.}~\bibnamefont{{Sandri}}},
  \bibinfo{journal}{Annals of Physics} \textbf{\bibinfo{volume}{24}},
  \bibinfo{pages}{332} (\bibinfo{year}{1963}{\natexlab{a}}).

\bibitem[{\citenamefont{{Sandri}}(1963{\natexlab{b}})}]{sandri2}
\bibinfo{author}{\bibfnamefont{G.}~\bibnamefont{{Sandri}}},
  \bibinfo{journal}{Annals of Physics} \textbf{\bibinfo{volume}{24}},
  \bibinfo{pages}{380} (\bibinfo{year}{1963}{\natexlab{b}}).

\bibitem[{\citenamefont{Chapman and Cowling}(1970)}]{livro_chapman}
\bibinfo{author}{\bibfnamefont{S.}~\bibnamefont{Chapman}} \bibnamefont{and}
  \bibinfo{author}{\bibfnamefont{T.}~\bibnamefont{Cowling}},
  \emph{\bibinfo{title}{The Mathematical Theory of Non-Uniform Gases}}
  (\bibinfo{publisher}{Cambridge University Press},
  \bibinfo{address}{Cambridge}, \bibinfo{year}{1970}), chap.
  \bibinfo{chapter}{The non-uniform state for a simple gas}.

\bibitem[{\citenamefont{{Albers} et~al.}(1971)\citenamefont{{Albers}, {Deutch},
  and {Oppenheim}}}]{1971_JChemPhys_54_3541}
\bibinfo{author}{\bibfnamefont{J.}~\bibnamefont{{Albers}}},
  \bibinfo{author}{\bibfnamefont{J.~M.} \bibnamefont{{Deutch}}},
  \bibnamefont{and}
  \bibinfo{author}{\bibfnamefont{I.}~\bibnamefont{{Oppenheim}}},
  \bibinfo{journal}{Journal of Chemical Physics} \textbf{\bibinfo{volume}{54}},
  \bibinfo{pages}{3541} (\bibinfo{year}{1971}).

\bibitem[{\citenamefont{van Kampen}(1992)}]{livro_vankampen}
\bibinfo{author}{\bibfnamefont{N.~C.} \bibnamefont{van Kampen}},
  \emph{\bibinfo{title}{Stochastic processes in Physics and Chemistry}}
  (\bibinfo{publisher}{North-Holland}, \bibinfo{address}{Amsterdam},
  \bibinfo{year}{1992}).

\bibitem[{\citenamefont{{Alder} and {Wainright}}(1970)}]{1970_PhysRevA_1_18}
\bibinfo{author}{\bibfnamefont{B.~J.} \bibnamefont{{Alder}}} \bibnamefont{and}
  \bibinfo{author}{\bibfnamefont{T.~E.} \bibnamefont{{Wainright}}},
  \bibinfo{journal}{Physical Review A} \textbf{\bibinfo{volume}{1}},
  \bibinfo{pages}{18} (\bibinfo{year}{1970}).

\bibitem[{\citenamefont{{Alder} et~al.}(1970)\citenamefont{{Alder}, {Gass}, and
  {Wainright}}}]{1970_JChemPhys_53_3813}
\bibinfo{author}{\bibfnamefont{B.~J.} \bibnamefont{{Alder}}},
  \bibinfo{author}{\bibfnamefont{D.~M.} \bibnamefont{{Gass}}},
  \bibnamefont{and} \bibinfo{author}{\bibfnamefont{T.~E.}
  \bibnamefont{{Wainright}}}, \bibinfo{journal}{Journal of Chemical Physics}
  \textbf{\bibinfo{volume}{53}}, \bibinfo{pages}{3813} (\bibinfo{year}{1970}).

\bibitem[{\citenamefont{{Duplantier}}(2007)}]{2007_arXiv_0705.1951}
\bibinfo{author}{\bibfnamefont{B.}~\bibnamefont{{Duplantier}}},
  \bibinfo{journal}{ArXiv e-prints}  (\bibinfo{year}{2007}),
  \eprint{arXiv:0705.1951}.

\bibitem[{\citenamefont{Nelson}(2006, 2nd Ed.)}]{livro_nelson}
\bibinfo{author}{\bibfnamefont{E.}~\bibnamefont{Nelson}},
  \emph{\bibinfo{title}{Dynamical Theories of Brownian Motion}}
  (\bibinfo{publisher}{Princeton University Press},
  \bibinfo{address}{Princeton}, \bibinfo{year}{2006, 2nd Ed.}),
  \bibinfo{note}{also available at:
  \texttt{www.math.princeton.edu/$\sim$nelson/books.html}}.

\bibitem[{\citenamefont{{Allahverdyan}
  et~al.}(2000)\citenamefont{{Allahverdyan}, {Nieuwenhuizen}, and
  {Saakian}}}]{2000_EPJB_16_317}
\bibinfo{author}{\bibfnamefont{A.~E.} \bibnamefont{{Allahverdyan}}},
  \bibinfo{author}{\bibfnamefont{T.~M.} \bibnamefont{{Nieuwenhuizen}}},
  \bibnamefont{and} \bibinfo{author}{\bibfnamefont{D.~B.}
  \bibnamefont{{Saakian}}}, \bibinfo{journal}{European Physical Journal B}
  \textbf{\bibinfo{volume}{16}}, \bibinfo{pages}{317} (\bibinfo{year}{2000}),
  \eprint{arXiv:cond-mat/9907090}.

\bibitem[{\citenamefont{{Kurchan}}(2005)}]{kurchan_nature}
\bibinfo{author}{\bibfnamefont{J.}~\bibnamefont{{Kurchan}}},
  \bibinfo{journal}{Nature} \textbf{\bibinfo{volume}{433}},
  \bibinfo{pages}{222} (\bibinfo{year}{2005}).

\bibitem[{\citenamefont{{Jung} and {Hanggi}}(1995)}]{1995_AdvChemPhys_89_239}
\bibinfo{author}{\bibfnamefont{P.}~\bibnamefont{{Jung}}} \bibnamefont{and}
  \bibinfo{author}{\bibfnamefont{P.}~\bibnamefont{{Hanggi}}},
  \bibinfo{journal}{Advances in Chemical Physics}
  \textbf{\bibinfo{volume}{89}}, \bibinfo{pages}{239} (\bibinfo{year}{1995}).

\bibitem[{\citenamefont{{Soares-Pinto} and
  {Morgado}}(2006)}]{2006_PhysA_365_289}
\bibinfo{author}{\bibfnamefont{D.~O.} \bibnamefont{{Soares-Pinto}}}
  \bibnamefont{and} \bibinfo{author}{\bibfnamefont{W.~A.~M.}
  \bibnamefont{{Morgado}}}, \bibinfo{journal}{Physica A}
  \textbf{\bibinfo{volume}{365}}, \bibinfo{pages}{289} (\bibinfo{year}{2006}),
  \eprint{arXiv:cond-mat/0601419}.

\bibitem[{\citenamefont{{Plyukhin}}(2006)}]{2006_EL_75_15}
\bibinfo{author}{\bibfnamefont{A.~V.} \bibnamefont{{Plyukhin}}},
  \bibinfo{journal}{Europhysics Letters} \textbf{\bibinfo{volume}{75}},
  \bibinfo{pages}{15} (\bibinfo{year}{2006}), \eprint{arXiv:cond-mat/0607005}.

\bibitem[{\citenamefont{{Dhar} and {Wagh}}(2007)}]{2007_EL_79_60003}
\bibinfo{author}{\bibfnamefont{A.}~\bibnamefont{{Dhar}}} \bibnamefont{and}
  \bibinfo{author}{\bibfnamefont{K.}~\bibnamefont{{Wagh}}},
  \bibinfo{journal}{Europhysics Letters} \textbf{\bibinfo{volume}{79}},
  \bibinfo{pages}{60003} (\bibinfo{year}{2007}),
  \eprint{arXiv:cond-mat/0604170}.

\bibitem[{\citenamefont{{Shokef} et~al.}(2007)\citenamefont{{Shokef},
  {Shulkind}, and {Levine}}}]{2007_PRE_76_030101}
\bibinfo{author}{\bibfnamefont{Y.}~\bibnamefont{{Shokef}}},
  \bibinfo{author}{\bibfnamefont{G.}~\bibnamefont{{Shulkind}}},
  \bibnamefont{and} \bibinfo{author}{\bibfnamefont{D.}~\bibnamefont{{Levine}}},
  \bibinfo{journal}{Physical Review E} \textbf{\bibinfo{volume}{76}},
  \bibinfo{pages}{030101} (\bibinfo{year}{2007}),
  \eprint{arXiv:cond-mat/0703040}.

\bibitem[{\citenamefont{{C{\'a}ceres}}(1999)}]{1999_PRE_60_5208}
\bibinfo{author}{\bibfnamefont{M.~O.} \bibnamefont{{C{\'a}ceres}}},
  \bibinfo{journal}{Physical Review E} \textbf{\bibinfo{volume}{60}},
  \bibinfo{pages}{5208} (\bibinfo{year}{1999}).

\bibitem[{\citenamefont{{Sancho} et~al.}(1982)\citenamefont{{Sancho}, {San
  Miguel}, {Katz}, and {Gunton}}}]{1982_PRA_26_1589}
\bibinfo{author}{\bibfnamefont{J.~M.} \bibnamefont{{Sancho}}},
  \bibinfo{author}{\bibfnamefont{M.}~\bibnamefont{{San Miguel}}},
  \bibinfo{author}{\bibfnamefont{S.~L.} \bibnamefont{{Katz}}},
  \bibnamefont{and} \bibinfo{author}{\bibfnamefont{J.~D.}
  \bibnamefont{{Gunton}}}, \bibinfo{journal}{Physical Review A}
  \textbf{\bibinfo{volume}{26}}, \bibinfo{pages}{1589} (\bibinfo{year}{1982}).

\bibitem[{\citenamefont{{Wio} et~al.}(1989)\citenamefont{{Wio}, {Colet}, {San
  Miguel}, {Pesquera}, and {Rodr{\'{\i}}guez}}}]{1989_PRA_40_7312}
\bibinfo{author}{\bibfnamefont{H.~S.} \bibnamefont{{Wio}}},
  \bibinfo{author}{\bibfnamefont{P.}~\bibnamefont{{Colet}}},
  \bibinfo{author}{\bibfnamefont{M.}~\bibnamefont{{San Miguel}}},
  \bibinfo{author}{\bibfnamefont{L.}~\bibnamefont{{Pesquera}}},
  \bibnamefont{and} \bibinfo{author}{\bibfnamefont{M.~A.}
  \bibnamefont{{Rodr{\'{\i}}guez}}}, \bibinfo{journal}{Physical Review A}
  \textbf{\bibinfo{volume}{40}}, \bibinfo{pages}{7312} (\bibinfo{year}{1989}).

\end{thebibliography}
\end{document}